\documentclass{article}
\usepackage{spconf,amsmath,graphicx}
\usepackage{amsmath}
\usepackage{amssymb}
\usepackage{array}
\usepackage{fixmath}
\usepackage{graphicx}
\usepackage{cite}
\usepackage{color}
\usepackage{changes}
\usepackage{algorithm}
\usepackage{algorithmic}
\usepackage{acronym}
\usepackage{subcaption}
\usepackage{bm}

\usepackage{mathtools}
\usepackage{cuted}
\usepackage{multirow}

\DeclareMathAlphabet{\mathbit}{OML}{cmr}{bx}{it}

\newacro{TDD}{time-division duplexing}
\newacro{CSI}{channel state information}
\newacro{DL}{downlink}
\newacro{UL}{uplink}
\newacro{BS}{base station}
\newacro{MS}{mobile station}
\newacro{MSE}{mean square error}
\newacro{MMSE}{minimum mean square error}
\newacro{SVD}{singular value decomposition}
\newacro{AM}{alternating minimization}
\newacro{OFDM}{orthogonal frequency-division multiplexing}
\newacro{mmWave}{millimeter wave}
\newacro{OMP}{orthogonal matching pursuit}
\newacro{MIMO}{multiple-input multiple-output}
\newacro{RF}{radio-frequency}
\newacro{LS}{least squares}
\newacro{MRC}{maximum ratio combiner}
\newacro{ZF}{zero-forcing}
\newacro{CS}{compressive sensing}
\newacro{ULA}{uniform linear array}
\newacro{ADC}{analog-to-digital converter}
\newacro{AoA}{angles-of-arrival}
\newacro{AoD}{angles-of-departure}
\newacro{CRLB}{Cram\'{e}r-Rao lower bound} 
\newacro{NMSE}{normalized mean squared error}
\newacro{CB}{coordinated beamforming}
\newacro{NMSE}{normalized mean-squared error}
\newacro{SINR}{signal-to-interference-plus-noise ratio}
\newacro{LLF}{log-likelihood function}
\newacro{SW-OMP}{simultaneous weighted - orthogonal matching pursuit}
\newacro{SS-SW-OMP+Th}{subcarrier-selection simultaneous weighted - orthogonal matching pursuit + thresholding}

\begin{document}

\title{
Separable multidimensional orthogonal matching pursuit and its application to joint localization and communication at mmWave} 
\name{Joan Palacios and Nuria Gonz\'{a}lez-Prelcic.}
\address{Electrical and Computer Engineering Department, North Carolina State University\\Email:\{\texttt{jbeltra,ngprelcic}\}\texttt{@ncsu.edu}}

\maketitle

\begin{abstract}
Greedy sparse recovery has become a popular tool in many applications, although its complexity  is still prohibitive when large sparsifying dictionaries or sensing matrices have to be exploited.
In this paper, we formulate first a new class of sparse recovery problems that exploit multidimensional dictionaries and the separability of the measurement matrices that appear in certain problems. Then we develop a new algorithm,  Separable Multidimensional Orthogonal Matching Pursuit (SMOMP),  which can solve this class of problems with low complexity. Finally, we apply SMOMP to the problem of joint localization and communication at mmWave, and numerically show its effectiveness to provide, at a reasonable complexity,  high accuracy channel and position estimations.
\end{abstract}

\section{Introduction}
Greedy algorithms for sparse recovery have become a popular tool for the reconstruction of the sparse signals 
that appear in applications such as compressive imaging, spectrum sensing or channel estimation for massive MIMO or millimeter wave (mmWave) MIMO \cite{Tropp2004}. In particular, orthogonal matching pursuit (OMP) \cite{Pati93orthogonalmatching,10.1117/12.173207} ] and simultaneous orthogonal matching pursuit (SOMP) \cite{Tropp2005,TROPP2006572}, 
have been extensively exploited in many applications where low complexity is desired.

The problem of channel estimation at mmWave exploiting a hybrid MIMO architecture, and more recently the problem of joint localization and communication, have been formulated and solved exploiting OMP, SOMP or some related variations of this algorithmic approach \cite{lee2014exploiting,Roi2015,Gao2016CL,Venugopal2017, SWOMP2018,Wu2019,Shahmansoori2018}. These strategies suffer, however,  from complexity limitations that come from two different aspects: 1) the exploitation of a very large sparsifying dictionary built from Kronecker products of the dictionaries for the direction of departure (DoD), the direction of arrival (DoA) and the delay; 3)
the large dimensionality of the measurement matrix representing the channel sounding process, which comes from the large number of antennas exploited at mmWave.

The Multidimensional Orthogonal Matching Pursuit (MOMP) algorithm \cite{MOMP} has been recently proposed to reduce the complexity of greedy approaches based on high resolution dictionaries built from Kronecker products. MOMP splits the projection step of  OMP into multiple, much simpler iterations, so it can be performed for each dictionary separately. 
Nonetheless, the MOMP formulation does not tackle the problem of having a large measurement matrix when the signal representation has a high dimensionality. For example, when applying MOMP to joint localization and communication at mmWave \cite{MOMP,Palacios2022Eusipco}, 
increasing the number of antennas or frequency carriers heavily increases the dimension of the measurement matrix, so it is barely possible to store it in an average computer memory or run the algorithm.

In this paper, we propose first an alternative algorithm to MOMP which does not require the explicit construction of the measurement matrix. Our new approach, Separable Multidimensional Orthogonal Matching Pursuit (SMOMP), builds upon the MOMP formulation, not only simplifying the computational complexity of the projection step when the number of dictionary elements increases, but also reducing complexity when the measurement matrix is separable. 
Then, we formulate the problem of joint  localization and communication at mmWave as a sparse reconstruction problem with multidimensional dictionaries and multiple measurement matrices, such that it can be solved with SMOMP. Numerical results show how SMOMP can provide channel estimates that lead to high accuracy positioning and high spectral efficiency in practical setups for mmWave MIMO where complexity and memory requirements prevent the execution of other greedy approaches.

Throughout the paper, $x$, ${\bf x}$, ${\bf X}$ and $\mathcal{X}$ will be the styles for scalar, vector, matrix or tensor and set. For a matrix ${\bf X}$, $[{\bf X}]_{a, :}$ and $[{\bf X}]_{:, b}$ are respectively, the $a$-th row and the $b$-th column, this notation is extended to tensors with multi-index like $[{\bf X}]_{{\bf a}, b} = [{\bf X}]_{a_1, a_2, b}$ for ${\bf a} = [a_1, a_2]$. The operator $\|{\bf x}\|$, $\|{\bf X}\|$ to denote the Euclidean and Frobenius norms.

\vspace*{-3mm}
\section{Separable multidimensional orthogonal matching pursuit}\label{sec:SMOMP}
\subsection{Background}
MOMP  is a generalization of OMP when working with multiple independent dictionaries \cite{MOMP,Palacios2022Eusipco}. To define the sparse recovery problem to be solved with MOMP, let us assume first that we have $N_{\rm D}$ dictionaries, with the $k$-th dictionary $\overline{\bf \Psi}_{k}\in\mathbb{C}^{N_k^{\rm s}\times N_k^{\rm a}}$ consisting of $N_k^{\rm a}$ atoms in $\mathbb{C}^{N_k^{\rm s}}$. Then we define the coefficients of the sparse signal in the set of dictionaries, i.e. $\overline{\bf C}\in\mathbb{C}^{N_{1}^{\rm a}\times\ldots\times N_{N_{\rm D}}^{\rm a}\times N^{\rm m}}$. We also need to write the measurement matrix as a tensor, i.e.,  $\overline{\bf \Phi}\in\mathbb{C}^{N^{\rm q}\times N_{1}^{\rm s}\times\ldots\times N_{N_{\rm D}}^{\rm s}}$.
Finally we define the set of entry coordinate combinations $\overline{\mathcal{I}} = \{\overline{\bf i}=(\overline{i}_1, \ldots, \overline{i}_{N_{\rm D}})\in\mathbb{N}^{N_{\rm D}}\text{ s.t. }\overline{i}_k \leq N_k^{\rm s}\quad\forall k \leq N_{\rm D}\}$, and the set of dictionary index combinations $\overline{\mathcal{J}} = \{\overline{\bf j}=(\overline{j}_1, \ldots, \overline{j}_{N_{\rm D}})\in\mathbb{N}^{N_{\rm D}}\text{ s.t. }\overline{j}_{k} \leq N_k^{\rm a}\quad\forall k \leq N_{\rm D}\}$ to cycle over each dictionary atom entry index and dictionary atom index, respectively.
With all these definitions, the equivalent multidimensional matching projection problem can now be formulated as
\vspace*{-2mm}
\begin{equation}\label{eq:MOMP}
\min_{\overline{\bf C}}\left\|\overline{\bf O}-\sum_{\overline{\bf i}\in\overline{\mathcal{I}}}\sum_{\overline{\bf j}\in\overline{\mathcal{J}}}[\overline{\bf \Phi}]_{:, \overline{\bf i}}\left(\prod_{k = 1}^{N_{\rm D}}[\overline{\bf \Psi}_{k}]_{\overline{i}_k, \overline{j}_k}\right)[\overline{\bf C}]_{\overline{\bf j}, :}\right\|^2,
\end{equation}
where $\overline{\bf O}\in\mathbb{C}^{N^{\rm q}\times N^{\rm m}}$ is the observation matrix. In this version of the problem, the sparsity condition is applied to the set $\overline{\mathcal{C}}\subset\overline{\mathcal{J}}$ of index $\overline{\bf j}\in\overline{\mathcal{J}}$ such that $\|\overline{\bf C}_{\overline{\bf j}, :}\| >0$.
It is proved in \cite{MOMP} that the multi-dimensional matching pursuit problem is an extension of the original matching pursuit formulation with multiple dictionaries that can be solved with MOMP. 
\vspace*{-1mm}
\subsection{Problem statement}
The memory requirements and complexity of operating with the measurement tensor  $\overline{\bf \Phi}$ in \eqref{eq:MOMP} are high when the number of dictionaries $N_{\rm D}$, atom sizes $N_{k}^{\rm a}$ and observation length $N^{\rm q}$ increase, since the total number of elements in $\overline{\bf \Phi}$ is given by $N^{\rm q}\prod_{k=1}^{N_{\rm D}}N_{k}^{\rm a}$.
In some problems, it is possible, however, to split this tensor into multiple ones and create independent measurements. We will formulate now the matching pursuit problem with multidimensional dictionaries for the cases in which it is possible to separate $\overline{\bf \Phi}$ into multiple matrices.  
First, we will need to define an additional index $f\leq N_{\rm F}$ to cicle over the different ${\bf \Phi}_f$ tensors.
This means that for each ${\bf \Phi}_f$, we will need its own collection of $N_f^{\rm D}$ dictionaries, with the $k$-th dictionary ${\bf \Psi}_{f, k}\in\mathbb{C}^{N_{f, k}^{\rm s}\times N_{f, k}^{\rm a}}$ consisting of $N_{f, k}^{\rm a}$ atoms in $\mathbb{C}^{N_{f, k}^{\rm s}}$.
This results in ${\bf \Phi}_f\in\mathbb{C}^{N_f^{\rm q}\otimes_{k=1}^{N_f^{\rm D}} N_{f, k}^{\rm s}}$, and consequently, we need to define the observation as ${\bf O}\in\mathbb{C}^{\otimes_{f=1}^{N_{\rm F}} N_{f}^{\rm q}\times N^{\rm m}}$.
For simplicity, we define a set of observation index combinations $\mathcal{O} = \{{\bf o}=(o_1, \ldots, o_{N_{\rm F}})\in\mathbb{N}^{N_{\rm F}}\text{ s.t. }o_f \leq N_f^{\rm q}\quad\forall f \leq N_{\rm F}\}$.
We keep in a single tensor the coefficients of the sparse signal in the multiple sets of dictionaries and we denote it as ${\bf C}\in\mathbb{C}^{\otimes_{f=1}^{N_{\rm F}}\otimes_{k=1}^{N_f^{\rm D}} N_{f, k}^{\rm s}\times N^{\rm m}}$.
We define index sets $\mathcal{I}_f = \{{\bf i}_f=(i_{f, 1}, \ldots, i_{f, N_{\rm D}})\in\bigotimes_{k=1}^{N_f^{\rm D}}\mathbb{N}_{N_{f, k}^{\rm s}}\}$, $\mathcal{J}_f = \{{\bf j}_f=(j_{f, 1}, \ldots, j_{f, N_{\rm D}})\in\bigotimes_{k=1}^{N_f^{\rm D}}\mathbb{N}_{N_{f, k}^{\rm a}}\}$, $\mathcal{I} = \{{\bf i}=({\bf i}_1, \ldots, {\bf i}_{N_{\rm F}})\in\otimes_{f=1}^{N_{\rm F}}\mathcal{I}_f\}$ and $\mathcal{J} = \{{\bf j}=({\bf j}_1, \ldots, {\bf j}_{N_{\rm F}})\in\otimes_{f=1}^{N_{\rm F}}\mathcal{J}_f\}$.
The equivalent separable multidimensional  matching projection problem is now defined as the minimization over ${\bf C}$ of
\vspace*{-5mm}
\begin{equation}\label{eq:SMOMP}
\sum_{{\bf o}\in\mathcal{O}}\hspace*{-0.5mm}\left\|[{\bf O}]_{{\bf o}, :}\hspace*{-1mm} -\hspace*{-1mm} \sum_{{\bf i}\in\mathcal{I}}\sum_{{\bf j}\in\mathcal{J}}\hspace*{-1mm}\prod_{f=1}^{N_{\rm F}}\hspace*{-1mm}\left(\hspace*{-1mm}[{\bf \Phi}_f]_{o_f, {\bf i}_f}\hspace*{-1mm}\prod_{k = 1}^{N_f^{\rm D}}[{\bf \Psi}_{f, k}]_{i_{f, k}, j_{f, k}}\hspace*{-1.5mm}\right)\hspace*{-1mm}[{\bf C}]_{{\bf j}, :}\right\|^2.
\end{equation}
To prove the equivalence with the MOMP formulation, we  compress ${\bf o}$ as a simple index $\overline{o}=\sum_{f=1}^{N_{\rm F}}o_f\prod_{f'=1}^{f-1}N_f^{\rm m}$, and group the $f, k$ index into the linear index $\overline{k} = k+\sum_{f'=1}^{f-1}N_{f'}^{D}$.
The last indices to reshape are $\overline{\bf i}\in\overline{\mathcal{I}}$ and $\overline{\bf j}\in\overline{\mathcal{J}}$ as $\overline{i}_{\overline{k}} = {i}_{f, k}$ and  $\overline{j}_{\overline{k}} = {j}_{f, k}$.
We can also compute the parameters $N^{\rm q} = \prod_{f=1}^{N_{\rm F}}N_f^{\rm q}$, $N_{\rm D} = \sum_{f=1}^{N_{\rm F}}N_f^{\rm D}$, $N_{\overline{k}}^{\rm a} = N_{f, k}^{\rm a}$ and $N_{\overline{k}}^{\rm s} = N_{f, k}^{\rm s}$.
This allows us to define the equivalent parameters as $[\overline{\bf O}]_{\overline{o}, :}=[{\bf O}]_{{\bf o}, :}$, $[\overline{\bf \Phi}]_{\overline{o}, {\bf i}} = \prod_{f=1}^{N_{\rm F}}[{\bf \Phi}_f]_{o_f, {\bf i}_f}$ and $\overline{\bf \Psi}_{\overline{k}} = {\bf \Psi}_{f, k}$.
Under these definitions, eq.~\eqref{eq:SMOMP} can be rewritten as eq.~\eqref{eq:MOMP}, proving them to be equivalent.
\vspace*{-5mm}
\subsection{Separable multidimensional matching pursuit}
The goal of this section is to describe SMOMP, a modification of MOMP to  solve problem \eqref{eq:SMOMP}. 
To do so, we analyze the steps in MMOP and describe the necessary transformations to define the corresponding step in SMOMP.

Like every OMP algorithm, MOMP starts by initializating the residual observation to $\overline{\bf O}_{\rm res}\leftarrow\overline{\bf O}$. This translates into SMOMP initializing ${\bf O}_{\rm res}\leftarrow{\bf O}$.
Next, we have an iterative process with two steps, namely projection step and residual update step.
For MOMP, the projection step is divided into two sub-steps, projection initialization and refinement.
To simplify the notation, the MOMP formulation makes use of the definition $\overline{\bf O}_{\rm \Phi}\in\mathbb{C}^{N^{\rm m}\otimes_{\overline{k}=1}^{^{N_{\rm D}}} N_{\overline{k}}^{\rm s}}$ as $[\overline{\bf O}_{\rm \Phi}]_{:, \overline{\bf i}} = \overline{\bf O}_{\rm res}^{\rm H}[\overline{\bf \Phi}]_{:, \overline{\bf i}}$. Equivalently,  SMOMP uses the definition $[{\bf O}_{\rm \Phi}]_{:, {\bf i}} = \sum_{{\bf o}\in\mathcal{O}}[{\bf O}_{\rm res}]_{{\bf o}, :}^{\rm H}\prod_{f=1}^{N_{\rm F}}[{\bf \Phi}_f]_{o_f, {\bf i}_f}$.

\begin{table*}
\centering
\begin{minipage}{0.75\textwidth}
\begin{equation}\label{eq:initialization_MOMP}
\max_{\hat{\overline{j}}_{\overline{k}}}\left(\frac{\sum_{\overline{k}''\in\overline{\hat{\mathcal{E}}}}\sum_{\overline{i}_{\overline{k}''=1}}^{N_{k''}^{\rm s}}\|\sum_{\overline{k}'''\in\overline{\mathcal{E}}\cup\{k\}}\sum_{\overline{i}_{\overline{k}'''=1}}^{N_{k'''}^{\rm s}}[\overline{\bf O}_{\rm \Phi}]_{:, \overline{\bf i}}[\overline{\bf \Psi}_{\overline{k}}]_{\overline{i}_{\overline{k}}, \hat{\overline{j}}_{\overline{k}}}\prod_{\overline{k}'\in\overline{\mathcal{E}}}[\overline{\bf \Psi}_{\overline{k}'}]_{\overline{i}_{\overline{k}'}, \hat{\overline{j}}_{\overline{k}'}}\|^2}{\sum_{\overline{k}''\in\overline{\hat{\mathcal{E}}}}\sum_{\overline{i}_{\overline{k}''=1}}^{N_{k''}^{\rm s}}\|\sum_{\overline{k}'''\in\overline{\mathcal{E}}\cup\{k\}}\sum_{\overline{i}_{\overline{k}'''=1}}^{N_{k'''}^{\rm s}}[\overline{\bf \Phi}]_{:, \overline{\bf i}}[\overline{\bf \Psi}_{\overline{k}}]_{\overline{i}_{\overline{k}}, \hat{\overline{j}}_{\overline{k}}}\prod_{\overline{k}'\in\overline{\mathcal{E}}}[\overline{\bf \Psi}_{\overline{k}'}]_{\overline{i}_{\overline{k}'}, \hat{\overline{j}}_{\overline{k}'}}\|^2}\right),
\end{equation}
\medskip
\hrule
\end{minipage}
\end{table*}
For the projection initialization, MOMP iteratively solves for the different $\overline{k}$ the expression in \eqref{eq:initialization_MOMP}, with $\overline{\mathcal{E}}$ being the set of already estimated indices $\overline{k}'$, and $\overline{\hat{\mathcal{E}}}$ being the set of indices $\overline{k}'$ which have not been estimated yet, excluding index $k$.

To adapt \eqref{eq:initialization_MOMP} to SMOMP, we define $\mathcal{E}$ as the set of index pairs $(f', k')$ already estimated, $\hat{\mathcal{E}}$ as the set of index pair which have not been estimated yet, excluding the pair $(f, k)$ and $\dot{\mathcal{E}} = \mathcal{E}\cup\{(f, k)\}$ and their slices $\mathcal{E}_f = \{(f', k')\in\mathcal{E}\text{ s.t. }f' = f\}$, $\hat{\mathcal{E}}_f = \{(f', k')\in\hat{\mathcal{E}}\text{ s.t. }f' = f\}$ and $\dot{\mathcal{E}}_f = \{(f', k')\in\dot{\mathcal{E}}\text{ s.t. }f' = f\}$.
To simplify the formulation, we also define, for any given set of indexes $\mathcal{A}$, the set $\mathcal{I}_{\mathcal{A}} = \{{\bf i}_{\mathcal{A}}: {\bf i}\in\mathcal{I}\}$, that is, the set of possible values of ${\bf i}_{\mathcal{A}}$.
Using this definition and the expression of the linear variables, applying the distributive property and eliminating constant terms ,we get the expression of the SMOMP projection initialization as
\begin{equation}\label{eq:initialization_SMOMP}
\max_{\hat{j}_{f, k}}\frac{\displaystyle\sum_{{{\bf i}_{\hat{\mathcal{E}}}\in\mathcal{I}_{\hat{\mathcal{E}}}}}\|\sum_{{{\bf i}_{\dot{\mathcal{E}}}\in\mathcal{I}_{\dot{\mathcal{E}}}}}[{\bf O}_{\rm \Phi}]_{:, {\bf i}}\prod_{\mathclap{(f', k')\in\dot{\mathcal{E}}}}[{\bf \Psi}_{f', k'}]_{i_{f', k'}, \hat{j}_{f', k'}}\|^2}{\displaystyle\sum_{{\bf i}_{\hat{\mathcal{E}}_{f}}\in\mathcal{I}_{\hat{\mathcal{E}}_{f}}}\|\sum_{{{\bf i}_{\dot{\mathcal{E}}_{f}}\in\mathcal{I}_{\dot{\mathcal{E}}_{f}}}}[{\bf \Phi}_{f}]_{:, {\bf i}_{f}}\prod_{\mathclap{(f', k')\in\dot{\mathcal{E}}_{f}}}[{\bf \Psi}_{f', k'}]_{i_{f', k'}, \hat{j}_{f', k'}}\|^2}.
\end{equation}
The expression in \eqref{eq:initialization_SMOMP} has the same solution as \eqref{eq:initialization_MOMP}, with a lower complexity due to the simplifications in the denominator.

For the projection refinement,  MOMP iteratively solves the following problem for the different $\overline{k}$, assuming all other indexes estimations to be known:
\vspace*{-1mm}
\begin{equation}\label{eq:refinement_MOMP}
\max_{\overline{j}_{\overline{k}}}\frac{\|\sum_{\overline{\bf i}\in\overline{\mathcal{I}}}[\overline{\bf O}_{\rm \Phi}]_{:, \overline{\bf i}}[\overline{\bf \Psi}_{\overline{k}}]_{\overline{i}_{\overline{k}}, \hat{\overline{j}}_{\overline{k}}}\prod_{\substack{\overline{k}'=1\\\overline{k}'\neq \overline{k}}}^{N_{\rm D}}[\overline{\bf \Psi}_{\overline{k}'}]_{\overline{i}_{\overline{k}'}, \hat{\overline{j}}_{\overline{k}'}}\|^2}{\|\sum_{\overline{\bf i}\in\overline{\mathcal{I}}}[\overline{\bf \Phi}]_{:, \overline{\bf i}}[\overline{\bf \Psi}_{\overline{k}}]_{\overline{i}_{\overline{k}}, \hat{\overline{j}}_{\overline{k}}}\prod_{\substack{\overline{k}'=1\\\overline{k}'\neq \overline{k}}}^{N_{\rm D}}[\overline{\bf \Psi}_{\overline{k}'}]_{\overline{i}_{\overline{k}'}, \hat{\overline{j}}_{\overline{k}'}}\|^2}.
\end{equation}
Following the same steps that we used to transform \eqref{eq:initialization_MOMP} into \eqref{eq:initialization_SMOMP}, we obtain the SMOMP projection refinement as
\begin{equation}\label{eq:refinement_SMOMP}
\max_{\hat{j}_{f, k}}\left(\frac{\displaystyle\|\sum_{{\bf i}\in\mathcal{I}}[{\bf O}_{\rm \Phi}]_{:, {\bf i}}\prod_{f'=1}^{N_{\rm F}}\prod_{k'=1}^{N_{f'}^{\rm D}}[{\bf \Psi}_{f', k'}]_{i_{f', k'}, \hat{j}_{f', k'}}\|^2}{\displaystyle\|\sum_{{\bf i}_f\in\mathcal{I}_f}[{\bf \Phi}_{f}]_{:, {\bf i}_{f}}\prod_{k'=1}^{N_{f}^{\rm D}}[{\bf \Psi}_{f, k'}]_{i_{f, k'}, \hat{j}_{f, k'}}\|^2}\right).
\end{equation}
The expression in \eqref{eq:refinement_SMOMP}, is equivalent to \eqref{eq:refinement_MOMP}, with a much lower complexity due to the huge simplification of the calculations in the denominator.

For the residual update step, MOMP updates $\overline{\bf O}_{\rm res}$ as $\overline{\bf O}-[\overline{\overline{\bf \Phi\Psi}}]_{:, \overline{\overline{\mathcal{C}}}}[\overline{\bf C}]_{\overline{\mathcal{C}}, :}$, where $[\overline{\overline{\bf \Phi\Psi}}]_{:, \overline{\overline{\mathcal{C}}}}$ is obtained by stacking the columns $\sum_{\overline{\bf i}\in\overline{\mathcal{I}}}[\overline{\bf \Psi}]_{:, \overline{\bf i}}\prod_{\overline{k}=1}^{\sum_f^{N_{\rm F}}N_f^{\rm D}}[\overline{\bf \Psi}_{\overline{k}}]_{\overline{i}_{\overline{k}}, \hat{\overline{j}}_{\overline{k}}}$ for $\hat{\bf j}\in\mathcal{C}$, and $[\overline{\bf C}]_{\overline{\mathcal{C}}, :}$ is the solution to $\min_{[\overline{\bf C}]_{\overline{\mathcal{C}}, :}}\|\overline{\bf O}-[\overline{\overline{\bf \Phi\Psi}}]_{:, \overline{\overline{\mathcal{C}}}}[\overline{\bf C}]_{\overline{\mathcal{C}}, :}\|^2$. In the case of SMOMP,  the columns of $[\overline{\overline{\bf \Phi\Psi}}]_{\overline{o}, \overline{\overline{\mathcal{C}}}}$ can be written as  
$\prod_{f=1}^{N_{\rm F}}\sum_{{\bf i}_f\in\mathcal{I}_f}[\overline{\bf \Psi}_f]_{o_f, {\bf i}_f}\prod_{k=1}^{N_f^{\rm D}}[{\bf \Psi}_{f, k}]_{i_{f, k}, \hat{j}_{f, k}}$.
We can then retrieve $[{\bf O}_{\rm res}]_{{\bf o}, :} = [\overline{\bf O}_{\rm res}]_{\overline{o}, :}$ and $[{\bf C}]_{\mathcal{C}, :}=[\overline{\bf C}]_{\overline{\mathcal{C}}, :}$.
This concludes the description of the SMOMP algorithm. An implementation of SMOMP is available online \cite{CodeSMOMPCore}.

\section{SMOMP-based joint channel estimation and localization}
We consider a MIMO communication system operating at mmWave frequencies based on a hybrid architecture and uniform rectangular arrays (URA) at both endsof sizes $N_{\rm T} = N_{\rm T}^{\rm x}\times N_{\rm T}^{\rm y}$ and $N_{\rm R} = N_{\rm R}^{\rm x}\times N_{\rm R}^{\rm y}$ and using $M_{\rm T}$ and $M_{\rm R}$ RF-chains at the transmitter and receiver respectively.
We consider the transmission of $N_{\rm S} \leq M_{\rm T}$ streams.
For training purposes, we choose square digital precoders and combiners, therefore $N_{\rm S}=M_{\rm T}$.
During the link establishment phase, the transmitter sends $M=M_1M_2$ sequences of $Q$ training symbol vectors to the receiver.
The $m-$th training frame, with $m = m_1M_2+m_2$, sounds the channel with a hybrid precoder ${\bf F}_{m_2}={\bf F}_{m_2}^{\rm RF}{\bf F}_{m_2}^{\rm BB}$ and a hybrid combiner ${\bf W}_{m_1}={\bf W}_{m_1}^{\rm RF}{\bf W}_{m_1}^{\rm BB}$ with their digital counterparts ${\bf F}_{m_2}^{\rm RF}\in\mathbb{U}_{\rm T}^{N_{\rm T}\times M_{\rm T}}$, ${\bf W}_{m_1}^{\rm RF}\in\mathbb{U}_{\rm R}^{N_{\rm R}\times M_{\rm R}}$, ${\bf F}_{m_2}^{\rm BB} \in\mathbb{C}^{M_{\rm T}\times M_{\rm T}}$ and ${\bf W}_{m_1}^{\rm BB} \in\mathbb{C}^{M_{\rm R}\times M_{\rm R}}$, for $\mathbb{U}_{\rm T}$ and $\mathbb{U}_{\rm R}$ the sets of feasible analog precoder and combiner entries.
$D$ is the delay spread.
The training symbol matrix of length $Q$ and $D$ symbols of zero padding for the $m$-th frame is denoted as ${\bf S}_{m_2} \in \mathbb{C}^{M_{\rm t}\times (Q+D)}$.

The frequency selective mmWave channel is modeled using a geometric channel model with $L$ paths \cite{Venugopal2017}.
The $d$-th delay tap of the channel, for $d\leq D$, is represented as
\vspace*{-2mm}
\begin{equation}\label{eq:geometric_channel}
{\bf H}_{d} = \sum_{l = 1}^{L}\alpha_l{\bf a}_{\rm R}{\bf a}_{\rm T}^{\rm H} p((d-1)T_{\rm s}+\tau_0-\tau_l),
\end{equation}
where $\alpha_l \in\mathbb{C}$, $\tau_l \in \mathbb{R}$ and ${\boldsymbol \theta}_l, {\boldsymbol \phi}_l \in \{{\bf v}\in\mathbb{R}^3\text{ s.t. }\|{\bf v}\|=1\}$ are the complex gain, delay, direction of arrival (DoA), and direction of departure (DoD) for the $l$-th path, $p(t)$ is the band limited pulse shaping filter including the contributions of the transmitter and receiver, $\tau_0$ is the clock offset, and ${\bf a}_{\rm T}$ and ${\bf a}_{\rm R}$ denote the steering vectors for the transmitter and the receiver.

Because of the multiple array configurations exploited simultaneously with a hybrid architecture, the transmission of each training frame will generate a set of received signals ${\bf Y}_m\in\mathbb{C}^{M_{\rm R}\times Q}$, comprised of $M_{\rm R}$ combinations of the $M_{\rm T}$ pilot signal streams.
Considering a transmission power $P$, the expression of ${\bf Y}_m$ is
\vspace*{-2mm}
\begin{equation}\label{eq:measurement_block}
[{\bf Y}_m]_{:, q} \! = \! \! \sqrt{P}\sum_{d = 1}^D\! {\bf W}_{m_1}^{\rm H}{\bf H}_d{\bf F}_{m_2}[{\bf S}_{m_2}]_{:, q+D-d} \! + \! {\bf W}_{m_1}^{\rm H}[{\bf N}_{m}]_{:, q},
\end{equation}
for a noise matrix ${\bf N}_{m}\in\mathbb{C}^{N_{\rm R}\times Q}$ with independent identically distributed entries following a distribution $\mathcal{NC}(0, \sigma^2)$, being $\sigma^2$ the noise power.

As proven in \cite{MOMP,Palacios2022Eusipco}, the channel estimation problem corresponds to the MOMP problem with $N_{\rm D} = 5$, $N^{\rm m} = 1$, the observation given by
\vspace*{-2mm}
\begin{equation}
[\overline{\bf o}]_{mM_{\rm R}Q+m_{\rm R}Q+q} = [{\bf L}_{m_1}^{-1}{\bf Y}_m]_{m_{\rm R}, q},\label{eq:MOMP_o}\\
\end{equation}
the measurement matrix defined as 
\vspace*{-4mm}
\begin{multline}
[\overline{\bf \Phi}]_{mM_{\rm R}Q+m_{\rm R}Q + q, {\bf i}} =\\ \sqrt{P}[{\bf L}_{m_1}^{-1}{\bf W}_{m_1}^{\rm H}]_{m_{\rm R}, i_1N_{\rm R}^{\rm y}+i_2}[{\bf F}_{m_2}{\bf S}_{m_2}]_{i_3N_{\rm T}^{\rm y}+i_4, q+D-i_5},\label{eq:MOMP_Phi}
\end{multline}
and dictionaries $\overline{\bf \Psi}_{1}$, $\overline{\bf \Psi}_{2}$, $\overline{\bf \Psi}_{3}$, $\overline{\bf \Psi}_{4}$ and $\overline{\bf \Psi}_{5}$ defined by the evaluation of ${\bf a}_{\rm R}^{\rm x}(\overline\theta^{\rm x})$,  ${\bf a}_{\rm R}^{\rm y}(\overline\theta^{\rm y})$, ${\bf a}_{\rm T}^{\rm x}(\overline\phi^{\rm x})$,  ${\bf a}_{\rm T}^{\rm y}(\overline\phi^{\rm y})$ and ${\bf a}_{\rm D}(\overline{\tau})$ respectively, in discrete domains of size $N_{k}^{\rm e}$.

Next, we transform the MOMP formulation into a SMOMP formulation.
From \eqref{eq:MOMP_Phi}, we see that this is an easy task when splitting the measurement matrix using the identity $m = m_1M_2+m_2$.
The MOMP problem thus becomes
\vspace*{-2mm}
\begin{align}
[{\bf O}]_{m_1M_{\rm R}+m_{\rm R}, m_2Q+q} = [{\bf L}_{m_1}^{-1}{\bf Y}_m]_{m_{\rm R}, q},\\
[{\bf \Phi}_1]_{m_1M_{\rm R}+m_{\rm R}, {\bf i}_1} = \sqrt{P}[{\bf L}_{m_1}^{-1}{\bf W}_{m_1}^{\rm H}]_{m_{\rm R}, i_{1, 1}N_{\rm R}^{\rm y}+i_{1, 2}},\\
[{\bf \Phi}_2]_{m_2Q+q, {\bf i}_2} = [{\bf F}_{m_2}{\bf S}_{m_2}]_{i_{2, 1}N_{\rm T}^{\rm y}+i_{2, 2}, q+D-i_{2, 3}},\\
{\bf \Psi}_{1, 1} = \overline{\bf \Psi}_{1},\quad {\bf \Psi}_{1, 2} = \overline{\bf \Psi}_{2},\\
{\bf \Psi}_{2, 1} = \overline{\bf \Psi}_{3},\quad {\bf \Psi}_{2, 2} = \overline{\bf \Psi}_{4},\quad {\bf \Psi}_{2, 3} = \overline{\bf \Psi}_{5}.
\end{align}

From the estimated channel parameters after training, it is possible to design the hybrid precoders and combiners that maximize the spectral efficiency  \cite{} and establish the link.
It is also possible to use the estimation for localization \cite{Shahmansoori2018,Wymeersch2018,Zhu2019,Jiang2021}.
In the numerical simulations section we will evaluate the quality of our channel estimation approach using communication and localization metrics.

\vspace*{-4mm}
\section{Numerical results}
We consider the uplink of an indoor mmWave MIMO system with two possible definitions of the system parameters. For System I,   $N_{\rm T}^{\rm x} = N_{\rm T}^{\rm y} = M_{\rm T} = 4$ and $N_{\rm R}^{\rm x} = N_{\rm R}^{\rm y} = M_{\rm R} = 8$, while for System II,   $N_{\rm T}^{\rm x} = N_{\rm T}^{\rm y} = M_{\rm T} = 8$ and $N_{\rm R}^{\rm x} = N_{\rm R}^{\rm y} = M_{\rm R} = 16$.

We generate the channels using a ray tracing simulation of a home office scenario as described in \cite{MOMP}.
The user has a height of $1.3{\rm m}$ and moves along $218$ different locations connecting to the access point with highest gain.
The noise power is set to $\sigma^2 = -81{\rm dBm}$. The delay spread is $D = 64$ and $Q=64$ training symbols are used.
To simplify the analysis, we make use of dictionaries of size $N_{f, k}^{\rm a} = K_{\rm res}N_{f, k}^{\rm s}$ for $K_{\rm res}=512$.
We build the pilot signal as the first $M_{\rm T}$ rows of a $64\times 64$ Hadamard matrix with 64 and 32 zeros of padding before and after the pilot.
${\bf W}_{m_1}$ and ${\bf F}_{m_2}$ are created by dividing the matrices resulting from the Kronecker product of DFT matrices with sizes $N^{\rm x}$ and $N^{\rm y}$ into blocks of as many columns as RF-chains.

We summarize first the comparison between MOMP and SMOMP in terms of memory requirements and execution time. The size of the measurement matrix for MOMP is $Q\frac{N_{\rm R}N_{\rm T}}{M_{\rm T}} \times N_{\rm T}N_{\rm R}D$, which results in less than 200 million elements for System I, while for System II it requires 8 billion elements, equivalent to more 128 Gb of memory. Because of this, MOMP can only run with the parameters in System I. For System I, the average computation time is $9.5{\rm s}$  for MOMP, while it can be reduced to $0.2{\rm s}$ with SMOMP. Both algorithms provide similar performance, with an average angular error in the estimation of strongest path of $0.34^\circ$ when the transmit power is set to 20 dBm. For System II, the average computation time is $1.4{\rm s}$ for SMOMP.

\begin{figure}[h!]
\centering
\includegraphics[width = 0.7\columnwidth]{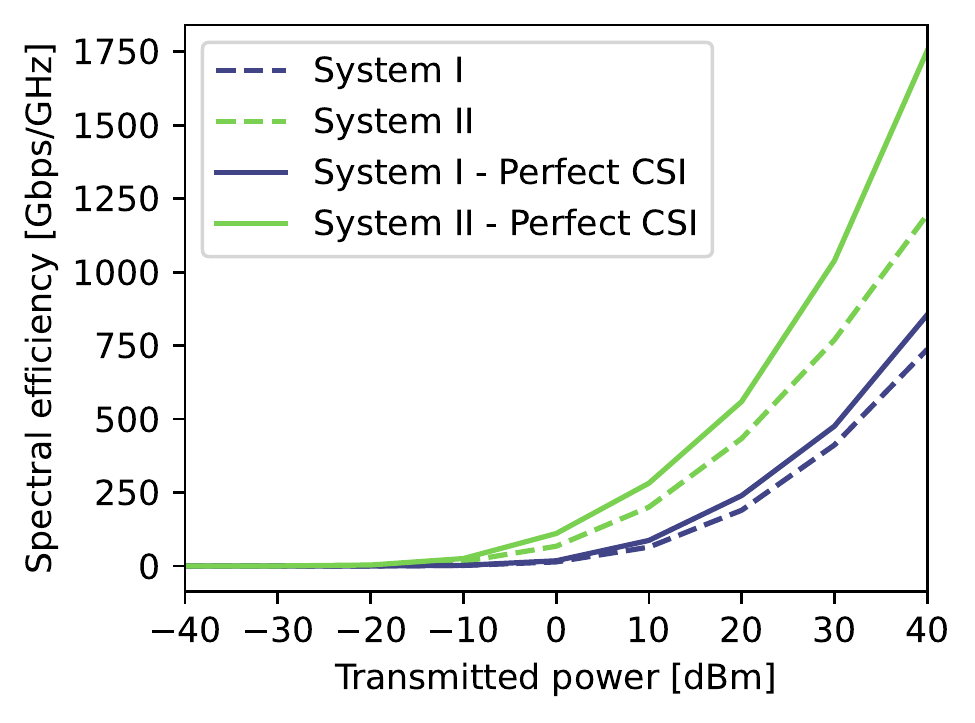}
\vspace*{-2mm}
\caption{Average spectral efficiency for System I and II.}
\label{fig:SE}
\end{figure}
\vspace*{-7mm}
We design the hybrid precoders and combiners through the approach PB in \cite{hybridref} and compute the spectral efficiency (SE) as defined in \cite{Venugopal2017}.
Fig.~\ref{fig:SE} shows the spectral efficiency provided by SMOMP as a function of the transmit power.
There is a small gap between the spectral efficiency achieved with hybrid precoders/combiners designed from the channel estimate and the one obtained with perfect knowledge.

\begin{figure}[h!]
\centering
\includegraphics[width = 0.7\columnwidth]{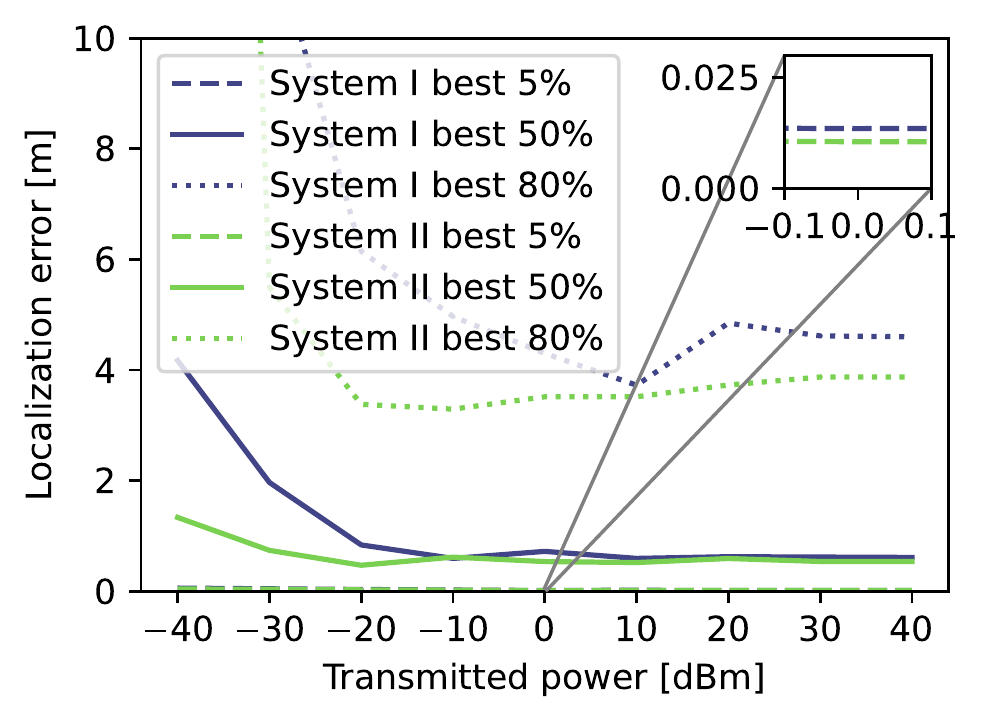}
\vspace*{-2mm}
\caption{Localization error as a function of the transmit power.}
\label{fig:loc}
\end{figure}
\vspace*{-2mm}
The localization error using the channel estimate and the positioning algorithm in \cite{MOMP} can be observed in Fig.~\ref{fig:loc} as a function of the transmit power.
For both System I and System II, over 50\% of the users can be located within sub-meter accuracy, while very high accuracy (in the order of cm) can be achieved for over 5\% of the users.  

\vspace*{-3mm}
\section{Conclusions}
We formulated a new class of sparse recovery problems with multidimensional dictionaries and multiple measurement matrices. We also proposed an algorithm called SMOMP to solve this type of problem with a reasonable complexity, which enables 
operation with high resolution dictionaries and large separable measurement matrices. 
We applied the proposed approach to the problem of joint channel estimation and localization at mmWave for a random deployment of users in an indoor scenario simulated by ray tracing.
We showed that the proposed approach can provide high accuracy results when considering practical system parameters where other greedy approaches pose unfeasible requirements in terms of memory and complexity.

\bibliographystyle{IEEEtran}  
\bibliography{refs,refs-MOMP}     


\end{document}